\documentclass[showpacs, aps, pre, twocolumn, groupedaddress, amsmath, amssymb]{revtex4}

\usepackage{graphicx}
\usepackage{dcolumn}
\usepackage{bm}

\begin{document}



\title{Effect of initial conditions on Glauber dynamics in complex networks}


\author{Makoto Uchida}
 \email{uchida@race.u-tokyo.ac.jp}
 \affiliation{Research into Artifacts, Center for Engineering (RACE), the University of Tokyo, 5-1-5 Kashiwanoha, Kashiwa, Chiba 277-8568, Japan}
\author{Susumu Shirayama}
 \email{sirayama@race.u-tokyo.ac.jp}
 \affiliation{Research into Artifacts, Center for Engineering (RACE), the University of Tokyo, 5-1-5 Kashiwanoha, Kashiwa, Chiba 277-8568, Japan}


\date{\today}

\begin{abstract}


The effect of initial spin configurations on zero-temperature Glauber
spin dynamics in complex networks is investigated.
In a system in which the initial spins are defined by centrality measures 
at the vertices of a network, a variety of non-trivial diffusive behaviors
arise, particularly in relation to functional relationships between 
the initial and final fractions of positive spins, 
some of which exhibit a critical point.
Notably, the majority spin in the initial state is not always dominant 
in the final state, and the phenomena that occur as a result of the dynamics
differ according on the initial condition, even for the same network.
It is thus concluded that the initial condition of a complex network exerts an 
influence on spin dynamics that is equally as strong as that exerted 
by the network structure.

\end{abstract}


\maketitle


The study of complex networks has generated broad interest, particularly in the field
of nonlinear physics \cite{Newman:2006, Boccaletti:2006}.
Dynamic processes and the dependence of such processes on networks are an important
topic in the field of complex networks, and many situations that result in non-trivial 
global phenomena have been discovered, despite the relative simplicity of 
local interactions. 
The Glauber dynamics of the Ising model, involving a set of spin-like binary variables 
and corresponding local interaction, are some of the simplest dynamics 
that occur in such networks. 
The Ising model with Glauber dynamics has been investigated as a model of the ordering 
process in spin glasses and other systems \cite{Bray:1994}. 
The interaction patterns of Glauber dynamics on regular {\itshape d}-dimensional lattices
and complex network structures have been considered in a number of studies, 
and it has been shown that such structures promote the development of characteristic 
ordering from a completely disordered state
\cite{Svenson:2001, Haeggstroem:2002, Boyer:2003, Castellano:2005, Castellano:2006}.

Yet beyond the effect of structural complexity itself, complex networks also exhibit
complexity related to the dynamics peculiar to complex structures. 
Previous of  Glauber dynamics have predominantly focused on the ordering dynamics 
that emerge from the completely disordered state, that is, the state in which the initial spins
are randomly distributed with an initial fraction $r$ of positive spins.
However, the vertices in complex networks are not interconnected uniformly,
instead they form a heterogeneous distribution of their degrees or locations. 
As a result of this heterogeneity, vertices in networks must be characterized topologically,
such as by the measure of {\itshape centrality} \cite{Freeman:1979},
which represents a set of parameters for individual vertices that are determined 
according to network topologies.
If vertices are discriminated by centrality measures, an initial configuration 
can be considered that is not randomly distributed, but rather topologically biased.
In such a case, it becomes possible to introduce some form of ordering into an initially
disordered state. Given initial conditions, Glauber dynamics can also be considered 
as controlling the diffusion of spins from arbitrarily selected vertices 
in the initial condition. 
Such a semi-ordered initial condition, in addition to the effect of the network structure,
is expected to have a significant effect on the resulting dynamics.
An attempt to demonstrate this effect through exploration of a simple two-state diffusion process
has been reported previously \cite{Uchida:2006}. However, while the numerical results can be
categorized into several classes, they are considered merely preliminary,
since the model is unfamiliar and numerical experiments alone are
insufficient to confirm such categorization.


In the present study, the dependence of zero-temperature Glauber dynamics, 
one of the simplest forms of Glauber dynamics, on the initial spin configuration
is investigated by Monte Carlo simulations of several types of complex networks.
The dynamics are taken from the arbitrary, semi-ordered initial spin configuration 
determined by the characteristics of the network structure.

Zero-temperature Glauber dynamics are considered here for the case of 
a complex network with spin variables $\sigma{}=\pm{}1$ located at
the vertices of the network.  The network is denoted by the adjacency matrix $A_{ij}$,
which is defined such that $A_{ij}=1$ if the vertices $i$ and $j$ are connected,
and $A_{ij}=0$ otherwise. In the present case of an undirected network, $A_{ij}$ is symmetric.
The local field $h_{i}(\tau{})$ acting on vertex $i$ during time step $\tau{}$ 
due to the spins of the neighboring vertices of the vertex $i$ is given by
\begin{equation}
h_{i}(\tau{}) = \sum_{j}^{N}A_{ij}\sigma_{j}(\tau{}) \quad (i=1...N).
\end{equation}
In the preceding study, similar dynamics were investigated as a two-state 
diffusion process \cite{Uchida:2006} by considering the following model: 
\begin{equation}
\sigma_{i}(n+1) = \begin{cases}
\operatorname{sgn} \left\{ h_{i}(n) \right\} \quad \text{if} \quad h_{i}(n) \ne 0 \\
\sigma(n) \quad \text{if} \quad h_{i}(n)=0. \\
\end{cases}
\end{equation}
In this model, the spins of vertices are updated synchronously at each time step $n$, 
that is, the spins at all vertices are updated simultaneously as $n$ progresses. 
In the present study, a general Monte Carlo method is employed for numerical simulations,
and the spins of individual vertices are updated asynchronously. 
The rule for updating the spin of a randomly selected vertex is as follows:
\begin{equation}
\sigma_{i}(\tau{}+1) = \begin{cases}
\operatorname{sgn} \left\{ h_{i}(\tau{}) \right\} \quad \text{if} \quad h_{i}(\tau{}) \ne 0 \\
\pm 1 \quad \text{(probability 1/2)} \quad \text{if} \quad h_{i}(\tau{})=0. \\
\end{cases}
\end{equation}
The progressive behavior of the spins is investigated using a normalized time step of $t = \tau{} / N$.

Spins on {\itshape d}-dimensional lattices have been investigated by a number of researchers,
and it has been shown that despite the simplicity of zero-temperature Glauber dynamics,
non-trivial phenomena arise under such regimes, even for regular lattices
\cite{Bray:1990, Sprin:2001, Sprin:2002}.
The Glauber dynamics of other types of complex networks, such as the Watts-Strogatz 
network \cite{Watts:1998}, have also been examined.
Boyer and Miramontes \cite{Boyer:2003}, by analyzing the ordering dynamics
of the Watts-Strogatz network, revealed a nonequilibrium ordering process 
induced by 'shortcuts' in the WS network, with occasional incomplete ordering.
Random graph network of Erd{\"o}s-R{\'e}nyi \cite{Erdos:1959} and scale-free networks
(i.e., Barab{\'a}si-Albert \cite{Barabasi:1999}) have also been investigated 
\cite{Castellano:2005}, for which Glauber dynamics do not lead to a fully ordered state.
In such systems, the dynamics may become trapped in a set of partially ordered 
(meta)stable states, even when the system is finite. 
Zhou and Lipowsky \cite{Zhou:2005} described a mean-field approach for analyzing
the dynamics in generic uncorrelated complex networks having arbitrary degree distributions.
In scale-free networks with $P(k) \sim{} k^{- \gamma{} }$, a characteristic power-law exponent
$\gamma{}_{c}$ at which a dynamic transition takes place was identified. 
The voter model, which is similar to zero-temperature Glauber dynamics, has also been studied
in complex networks \cite{Castellano:2003, Vilone:2004, Suchecki:2005, Sood:2005}.
Remarkably, some of these studies have indicated that the majority population 
in the initial state may not dominate asymptotically \cite{Sood:2005}.

In these previous studies, the initial spins of the network vertices are considered 
to be randomly distributed, and it remains to be discussed whether a degree of order
in the initial state affects the network dynamics. 
The measure of centrality has been used in network science to represent the characteristics
of vertices relative to other vertices \cite{Freeman:1979, Wasserman:1994}.
The use of centrality to characterize vertices thus allows a certain degree of order to be
introduced into the initial spin configuration to which Glauber dynamics are applied.
In the present study, the dependence of the dynamics on the initial spin configuration is
examined using three kinds of centrality measures: degree centrality $C_{deg}(v)$,
closeness centrality $C_{close}(v)$, and betweenness centrality $C_{bet}(v)$ \cite{Brandes:2001}.
The clustering coefficient $C_{clust}(v)$ is also considered as a characteristic of vertices
\cite{Watts:1998}. These measures are defined for a vertex $v$ as follows:
\begin{eqnarray}
C_{deg}(v) &=& \frac{k_{v}}{N-1} \\
C_{close}(v) &=& \frac{1}{\sum_{t}d_{G}(v, t)} \\
C_{bet}(v) &=& \sum_{s\ne{}t\ne{}v}\frac{\sigma_{st}(v)}{\sigma_{st}} \\
C_{clust}(v) &=& \frac{E}{
\left(
\begin{array}{c}
k_{v} \\
2
\end{array}
\right)
}
\end{eqnarray}
where $k_{v}$ is the degree of the vertex $v$, $d_{G}(v, t)$ is the distance between vertices $v$ and $t$,
$\sigma_{st}$ and $\sigma_{st}(v)$ are the number of shortest paths from $s$ to $t$ 
and from $s$ to $t$ via $v$, and $E$ is the number of edges between the neighbors of vertex $v$.

Corresponding to the centralities of each vertex, the $rN$ vertices with the largest centrality 
are assigned the positive spin state in the initial condition ($\sigma{}(0) = +1$),
while the remaining $(1-r)N$ vertices are assigned $\sigma{}(0) = -1$.
Here, $r$ is the initial fraction of positive spins, and $N$ is the number of vertices
in the network. 
Four types of complex network models are investigated using these initial conditions:
Random graph network of Erd{\"o}s-R{\'e}nyi \cite{Erdos:1959}, and three types of structured
scale-free networks -- the B{\'a}rabasi-Albert (BA) network \cite{Barabasi:1999},
the Klemm-Egu{\'i}lz (KE) network (also reffered to as the highly clustered scale-free network)
\cite{Klemm:2002b}, and the connecting nearest neighbor (CNN) network \cite{Vazquez:2003}.
In the highly clustered scale-free network, the mixing parameter is set to $\mu{} = 0.1$.
All the scale-free network models are categorized as evolving network models 
in which the distributions of the degree $k$ follow a power-law distribution
$P(k) \sim{} k^{- \gamma{} }$. 
However, the networks differ in that the BA network has a low clustering coefficient
$C \sim N^{-0.75}$ while the other two networks are much more highly clustered 
($C \sim \mathcal{O}(1)$), and the CNN network has a positive (assortativie) degree correlation.
See References \cite{Barabasi:1999, Klemm:2002b, Vazquez:2003} for further details.
The same network models were used with a fixed number of vertices and average
degrees in the preceding study \cite{Uchida:2006}. In the present paper,
the scaling effects of system size are also considered. 
In the series of simulations described below, all the vertices in the networks are
confirmed to be {\itshape connected} by edges so taht they form a single component.  

\begin{figure*}
  \centering

 \begin{tabular}{cc}

   \resizebox{0.48\linewidth}{!}{ \includegraphics[]{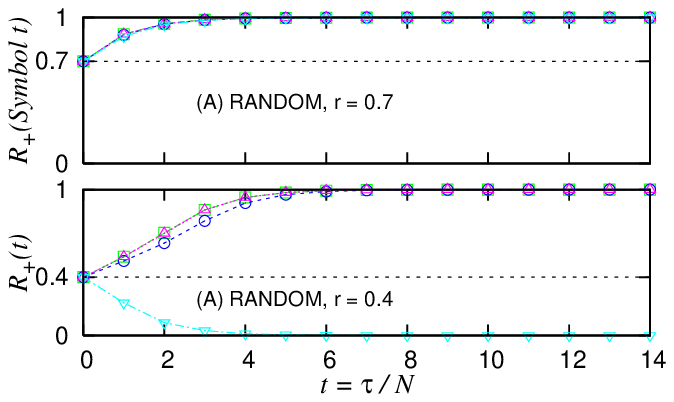} } &
   \resizebox{0.48\linewidth}{!}{ \includegraphics[]{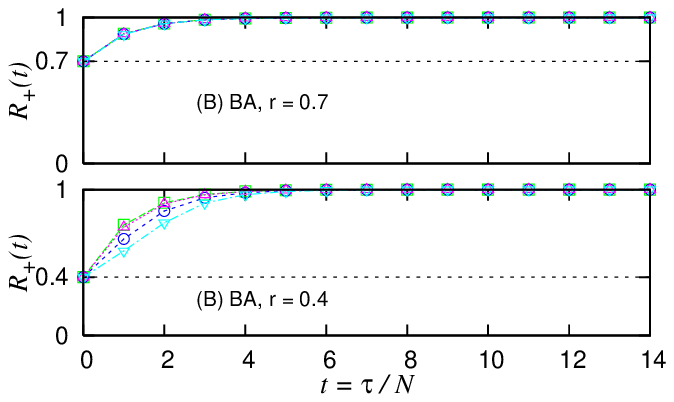} } \\
   \resizebox{0.48\linewidth}{!}{ \includegraphics[]{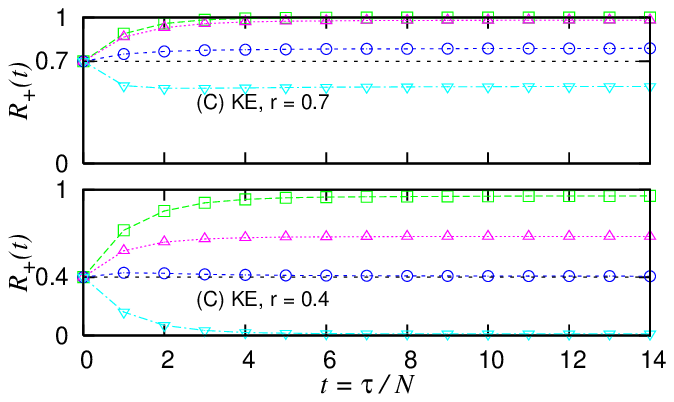} } &
   \resizebox{0.48\linewidth}{!}{ \includegraphics[]{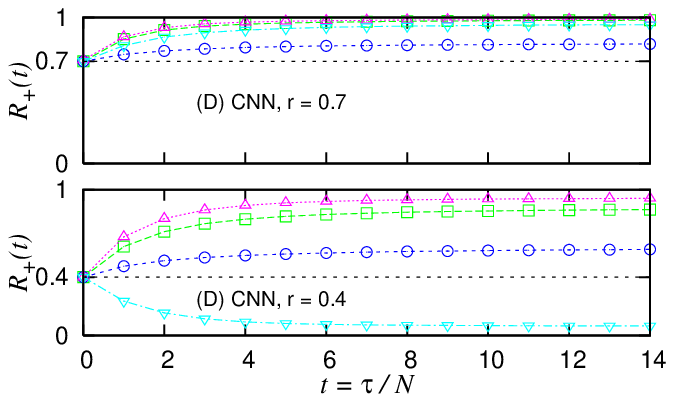} } \\
 \end{tabular}

 \caption{ \label{fig:te} Temporal evolution of $R_{+}(t)$ in a random network (A), a BA network (B), a KE network (C) and a CNN network (D). Initial fractions $r=0.4$ and $0.7$. The different symbols denote the ordering of the initial state, according to degree centrality ($\square{}$), closeness centrality ($\circ{}$), betweenness centrality ($\vartriangle{}$) and clustering coefficient ($\triangledown{}$). $N=18000$ and $\langle{} k \rangle{} = 10$. } 
\end{figure*}

\begin{figure*}
  \centering

 \begin{tabular}{cc}

   \resizebox{0.40\linewidth}{!}{ \includegraphics[]{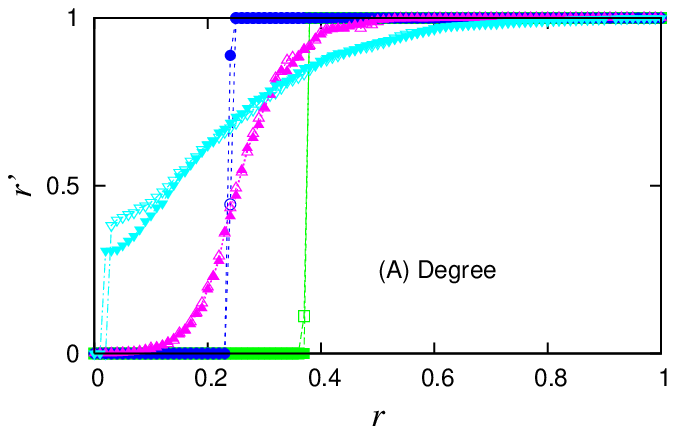} } &
   \resizebox{0.40\linewidth}{!}{ \includegraphics[]{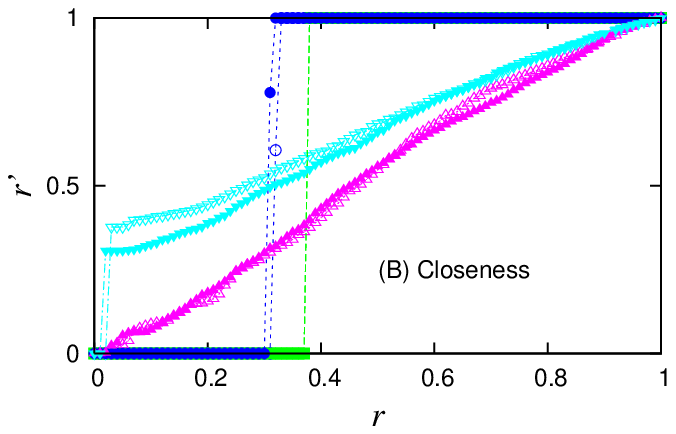} } \\
   \resizebox{0.40\linewidth}{!}{ \includegraphics[]{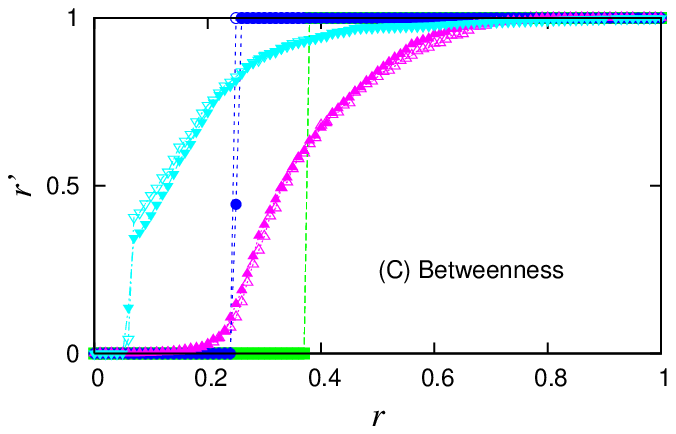} } &
   \resizebox{0.40\linewidth}{!}{ \includegraphics[]{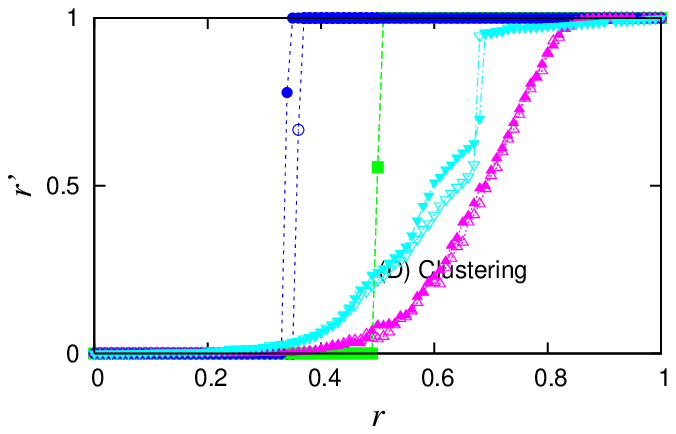} } \\
 \end{tabular}

  \caption{ \label{fig:conv_model1} Convergence fraction of positive spins $r'$ as a function of 
            initial fraction $r$, derived from the initial ordering by (A) degree centrality,
	    (B) closeness centrality, (C) betweenness centrality and (D) clustering coefficient, 
	    on the random network ($\square{}$), the BA network ($\circ{}$), the KE network ($\vartriangle{}$) 
	    and the CNN network ($\triangledown{}$).
        Open symbols and closed symbols denote data for system size $N = 18000$ and $N = 36000$,
        respectively.  $\langle{} k \rangle{} = 10$.
	    }
\end{figure*}

\begin{figure*}
  \centering

 \begin{tabular}{cc}

   \resizebox{0.40\linewidth}{!}{ \includegraphics[]{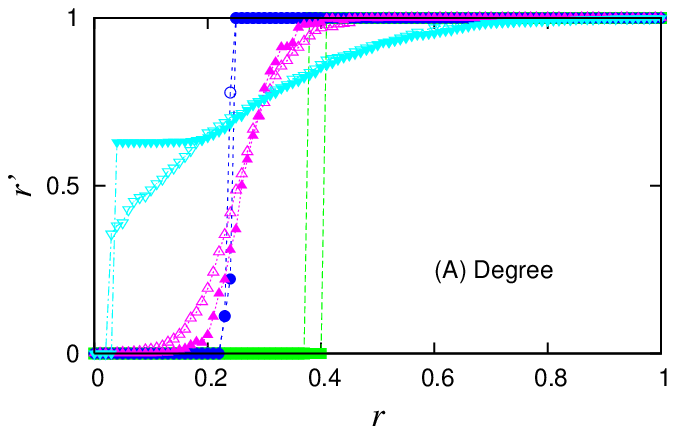} } &
   \resizebox{0.40\linewidth}{!}{ \includegraphics[]{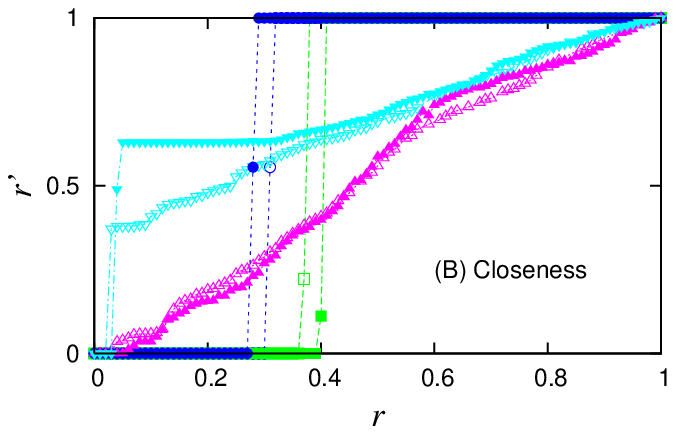} } \\
   \resizebox{0.40\linewidth}{!}{ \includegraphics[]{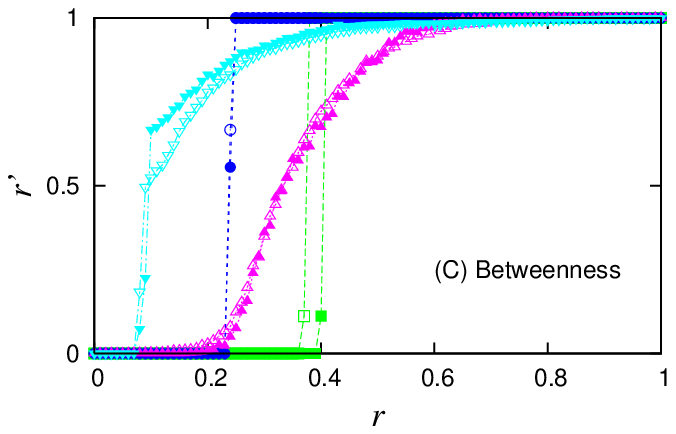} } &
   \resizebox{0.40\linewidth}{!}{ \includegraphics[]{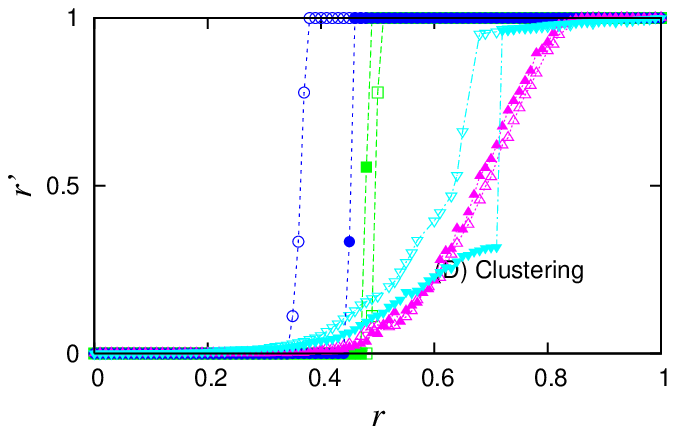} } \\
 \end{tabular}

  \caption{ \label{fig:conv_model2} Same quantities and symbols as Fig.~\ref{fig:conv_model1}.
            Open and closed symbols and filled symbols denote the plot for the average degree
	    $\langle{} k \rangle{} = 10$ and $\langle{} k \rangle{} = 18$, respectively. 
	    The system size is $N = 9000$. 
  }
\end{figure*}

\begin{figure}
   \resizebox{0.80\linewidth}{!}{ \includegraphics[]{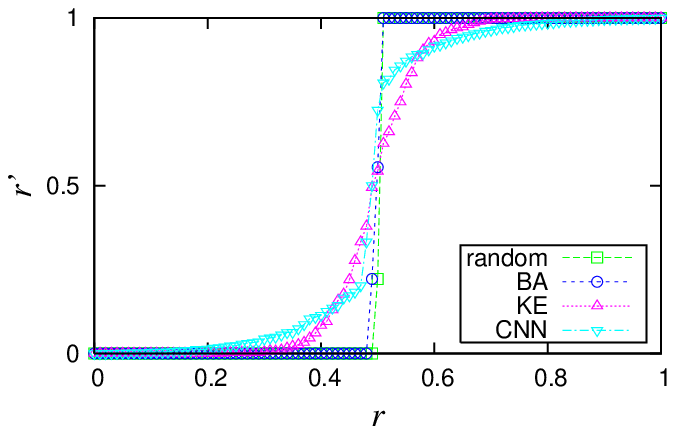} } 
   \caption{ \label{fig:random_cv} Convergence fraction of positive spins $r'$
             as a function of initial fraction $r$ derived from a completely disordered 
             (i.e., randomly distributed) spin state. $N=18000$, $\langle{} k \rangle{} = 10$.}
\end{figure}


The progression of positive spin fraction $R_{+}(t)$ at $t=\tau{}/N$, starting from fixed 
initial fractions of $r=0.4$ and $r=0.7$ and a variety of initial orderings of positive spins 
is shown in Fig.~\ref{fig:te}.
The results shown throughout this report are averages obtained over several simulations.
In the random and BA networks, the dynamics reach a fully ordered state in which all 
of the vertices share the same spin state. 
With $r=0.4$ in the initial state, positive spins derived from vertices having larger centralities,
despite being the minority, spread over the entire network in all cases except the BA network
with the initial ordaring based on the clustering coefficient measure.
In the KE and CNN networks, the dynamics reach a metastable state consisting of two coexisting
spin states with fraction $r'$, which differs from $r$.


Figures~\ref{fig:conv_model1} and \ref{fig:conv_model2} show the relationships between 
the initial fraction $r = R_{+}(0)$ and the convergence fraction $r' = R_{+}(\infty{})$
of positive spins in systems having various sizes and average degrees.
Several classes of characteristic functional forms can be observed. 
If the respective numbers of two spin states are not equal (i.e., $r \ne{} 0.5$), 
and the two states are randomly distributed, it is obvious that the state having
the larger quantity in the initial state will become dominant at $ t = \infty{} $. 
This was verified by performing preliminary simulations (Fig.~\ref{fig:random_cv}). 
In the random and BA networks, the convergence fraction of the two spin states exhibits
a step-like change with respect to initial conditions, where the critical fraction $r = r_{c} < 0.5$
is non-trivial. Notably, $r_{c}$ is always lower in the BA network than in the random network.


In contrast, the KE and CNN networks produce much more complex variation in the convergence
fraction of the two spin states with respect to ordering of the initial spin configuration,
reflecting the emergence of a metastable state. 
The initial states with ordering defined by degree and betweenness centrality exhibit convex 
or sigmoidal functions over $r$, whereas the relationship for initial ordering defined by
closeness centrality is almost linear. 
A critical point of $r = r_{c}$ can also be identified in these traces, around which $r'$
jumps or rapidly increases in some cases. 
If the initial spins are completely disordered, $r_{c}$ should take a value of $0.5$.
However, the value of $r_{c}$ differs according to the initial condition applied.

\begin{figure}
  \centering

 \begin{tabular}{c}

   \resizebox{0.80\linewidth}{!}{ \includegraphics[]{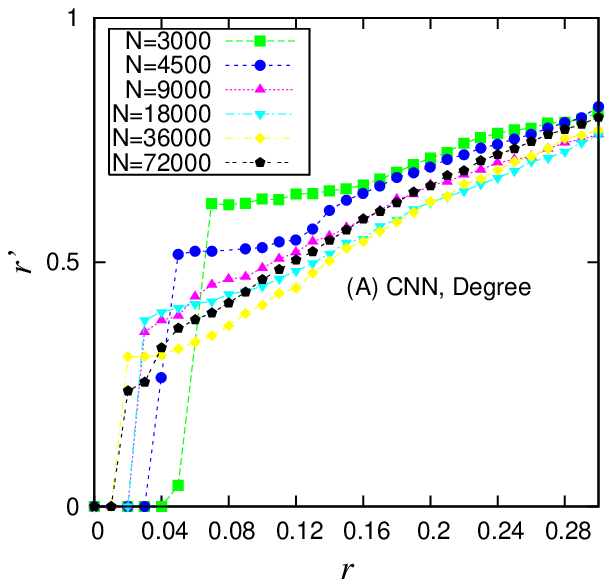} } \\
   \resizebox{0.80\linewidth}{!}{ \includegraphics[]{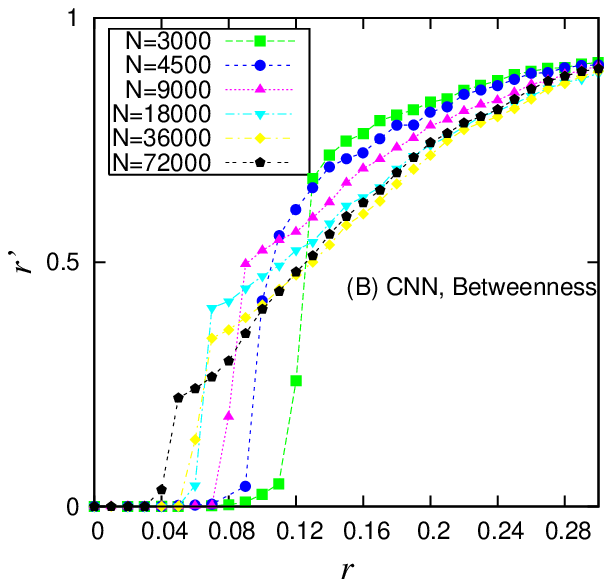} } \\
 \end{tabular}

  \caption { \label{fig:cnn_scale} Convergence fraction of positive spins $r'$ around             
             the critical point $r_{c}$ as a function of initial fraction $r$ for the CNN networks 
             of various sizes with initial orderings defined by (A) degree centrality and
             (B) betweenness centrality.
  }

\end{figure}
\begin{figure}
  \centering

 \begin{tabular}{cc}

   \resizebox{0.80\linewidth}{!}{ \includegraphics[]{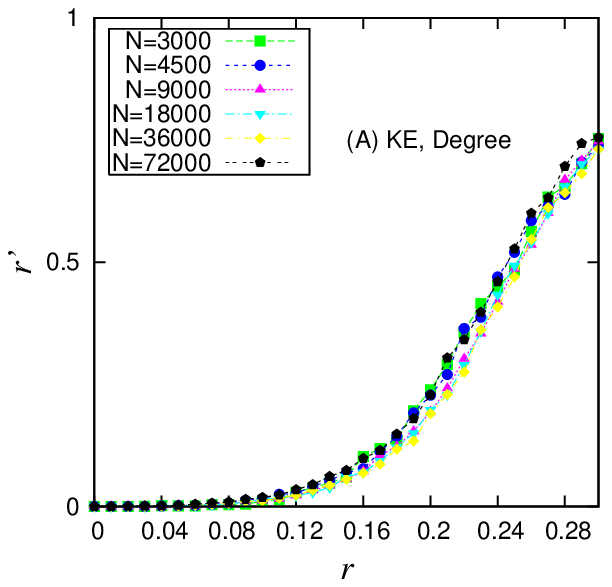} } \\
   \resizebox{0.80\linewidth}{!}{ \includegraphics[]{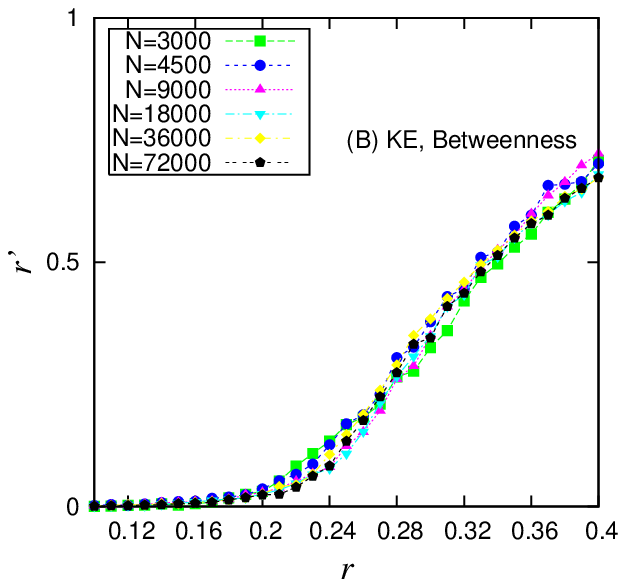} } \\
 \end{tabular}

  \caption { \label{fig:ke_scale} Same quantities and symbols as Fig.~\ref{fig:cnn_scale} for the KE networks.  
  }
\end{figure}

Figures~\ref{fig:conv_model1} and \ref{fig:conv_model2} show the $r$-$r'$
relationships for systems of various sizes and average degrees.
Although some differences can be seen in the detailed shape of the relationship 
and the value of $r_{c}$, comparison of Figs.~\ref{fig:conv_model1} and \ref{fig:conv_model2}
suggest that the class of the characteristic function is independent of the system size
and the average degrees, except in the case of the CNN network. 
To investigate the scaling effects in more detail, the critical behavior of the $r$-$r'$
relationships was examined over a wider range of system size considering a representative
set of initial orderings. A finite scaling on the CNN network can be recognized in both cases
shown in Fig.~\ref{fig:cnn_scale}. As the system size increases, the critical point $r_{c}$
asymptotically approaches a certain value, and the transition at $r_{c}$ is not abrupt.
No such scaling effects can be recognized on the KE network from Fig.~\ref{fig:ke_scale},
where the shape of the function and the value of $r_{c}$ remains relatively constant,
independent of system size.

%

\begin{figure*}
  \centering

 \begin{tabular}{cc}

   \resizebox{0.48\linewidth}{!}{ \includegraphics[]{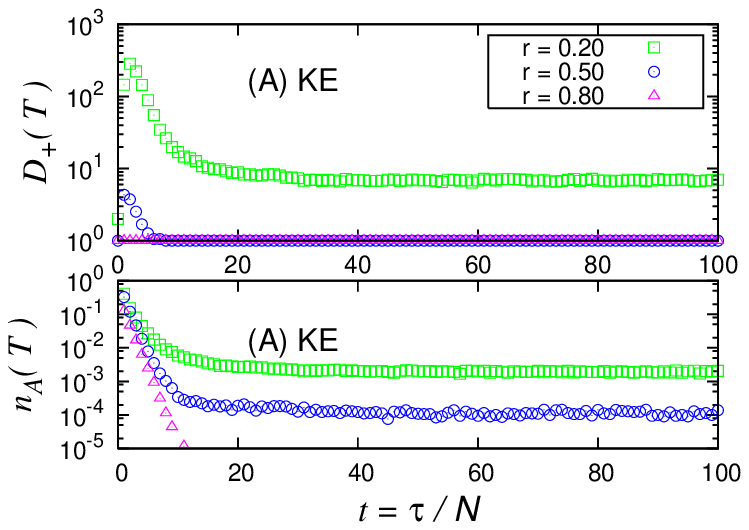} } &
   \resizebox{0.48\linewidth}{!}{ \includegraphics[]{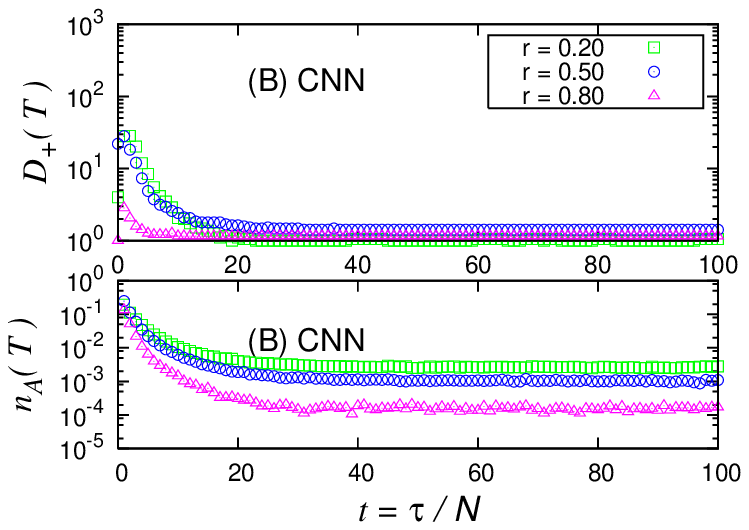} } \\
 \end{tabular}

  \caption{ \label{fig:ke_cnn_ad} Evolution of fraction of updating vertices 
            (i.e., vertices that change state) (lower), and number of connected domains 
            of positive spins (upper) on (A) KE and (B) CNN networks for $r=0.20$, $0.50$
            and  $0.80$. Initial spins ordered by degree centrality. $N=18000$, $\langle{} k \rangle{} = 10$.  }
\end{figure*}


The characteristics of the metastable state emerging in the KE and CNN networks are
illustrated using the fraction of vertices that flip spin (i.e., the vertices that
connect to an equal number of different spins, or the vertices located at 
the boundary of a domain of identical spins), and the number of connected domains
of positive spins at $t = \tau{}/N$.
These parameters are denoted by $n_{A}(t)$ and $D_{+}(t)$, respectively.
Figure~\ref{fig:ke_cnn_ad} shows an example of the variation in $n_{A}(t)$ and $D_{+}(t)$
with $r$ in systems with initial ordering defined by degree centrality. 
It can be seen that $D_{+}(t)$ does not always reach unity, suggesting that 
the metastable state can consist of several separated clusters of identical spins.
Moreover, $n_{A}(t)$ does not converge to zero, indicating that the metastable state is 
not static but rather a stationary active state. 
The results are essentially the same for other initial orderings. 
It has been reported that a similar phenomenon might occur in a random network from
a completely disordered initial state \cite{Castellano:2005}, implying that the final 
state consists of two large domains of opposite spin. 
A similar investigation was conducted using high-dimensional lattices \cite{Sprin:2002}.
The present simulations considering a range of complex network structures and arbitrary 
initial conditions suggest an alternative outcome is possible.

\begin{figure*}
  \centering

 \begin{tabular}{cc}
   \resizebox{0.48\linewidth}{!}{ \includegraphics[]{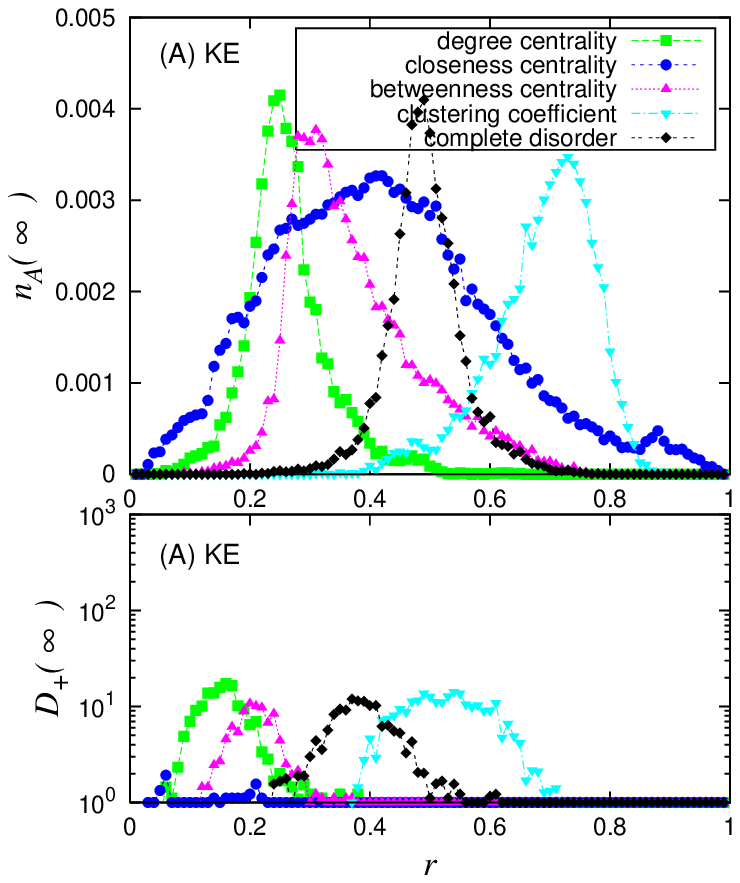} } &
   \resizebox{0.48\linewidth}{!}{ \includegraphics[]{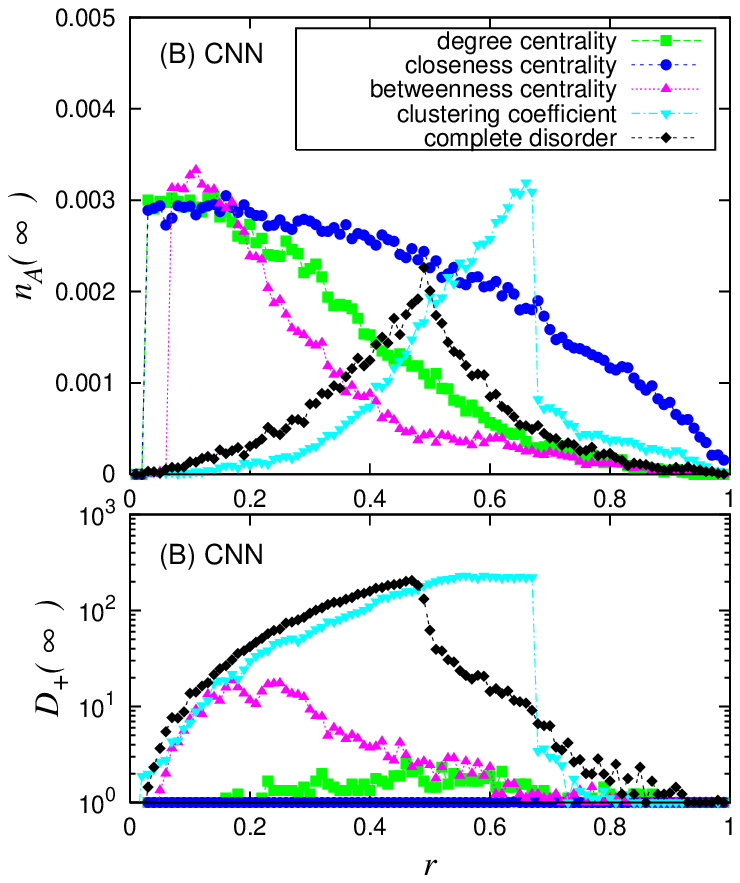} } \\
 \end{tabular}

        
  \caption{ \label{fig:ke_cnn_ad_int} Fraction of updating vertices, and number of connected 
            domains of positive spin in the metastable state (at $t = \infty{} $) with respect 
            to $r$ on (A) KE and (B) CNN networks considering various orderings of initial spins.
            $N=18000$, $\langle{} k \rangle{} = 10$. }

\end{figure*}

The plots of $n_{A}(t)$ and $D_{+}(t)$ in the final metastable state (at $t = \infty{}$) 
with respect to $r$ for various initial conditions (Fig.~\ref{fig:ke_cnn_ad_int}) indicate
that the final states for the two networks are quite different.
On the KE network, $n_{A}( \infty{} )$ exhibits a clear peak for each set of initial conditions,
whereas a more complicated noncontinuous function appears on the CNN network.
For a completely disordered initial configuration, $n_{A}( \infty{} )$ reaches a maximum
at $r = 0.5$ and forms a symmetric function.
For most initial conditions on both networks, the initial fraction $r$ affording the maximum 
$n_{A}( \infty{} )$ is consistent with the critical fraction $r=r_{c}$ in Fig.~\ref{fig:conv_model1}.
It is therefore safe to conclude that the metastable state arising from $r \simeq r_{c}$ 
is somewhat unstable.
The number of connected domains of positive spins (Fig.~\ref{fig:ke_cnn_ad_int}) reveals 
another characteristic. At small $r$, positive spins tend to shrink and become extinct, 
thus $D_{+}( \infty{} ) = 0$. As $r$ increases, several clusters of positive spin are
able to persist, causing $D_{+} ( \infty {} )$ to increase. 
As $r$ increases further, the fragmented domains begin to agglomerate to form fewer, 
larger clusters, until a singl large domain emerges (i.e., $D_{+}( \infty{} ) = 1$). 
The degree of fragmentation depends on the initial conditions and the networks. 
It should be noted that $D_{+}( \infty{} )$ is always equal to or very close to unity
in both networks when the order of the initial spin configuration is defined by 
the closeness centrality measure. 
It is suspected that this gives rise to the linear relationship between $r$ and $r'$,
as shown in Fig.~\ref{fig:conv_model1}(B).


In summary, the effect of the initial spin configuration on zero-temperature 
Glauber dynamics acting on various types of complex networks was examined. 
Through a series of numerical analyses it was revealed that non-trivial diffusive behavior
that depends on both the initial condition and the network structure occurs in such systems.
In some cases, the final dynamics reached a metastable state involving two coexistent 
spin states with several connected domains of identical spins, the marginal vertices of 
which flip continuously. 
The functional relationship between the initial fraction $r$ of positive spins and 
the final fraction $r'$ can be categorized into one of several patterns, some of which exhibit
critical point $r_{c}$. 
The patterns and criticality were found on networks of various system sizes and average
degrees. A finite scaling effect with respect to system size was identified 
on the CNN network, whereas the other networks examined exhibited no such scaling effect.

Although a clear understanding of the origin of the behavior revealed in this study has 
yet to be obtained, some observations can be made on the general characteristics of
these phenomena. As the KE and CNN networks are known to be more highly clustered than
the random and BA networks, the final metastable state may originate from the topological
cohesiveness of such networks.
However, the characteristic functional relationships between $r$ and $r'$,
and the origins of the critical point $r_{c}$, remain unclear. One possible feature
that could be responsible for these characteristics is the correlation in networks
\cite{Pastor-Satorras:2001b}. The CNN network is known to have a positive degree 
correlation \cite{Vazquez:2003}, and hence if the vertices of higher degree are 
more likely to be interconnected, the positive spins originating from these vertices
will not disappear, even for a small initial fraction.
For other initial orderings, the correlation of centralities may also play
a key role in the unique dynamics observed. 
The results of the present study may therefore lead to novel understandings of 
the underlying characteristics of individual vertices measured thus far by centralities. 
It is also expected that further characteristics of network topologies remain to be revealed. 
The nonequilibrium spin dynamics exhibited by these systems do not appear to be comprehensively 
explainable based on topological scaling or conventional measures of the underlying complex 
networks alone. Nevertheless, the present study has revealed a strong dependence of spin dynamics
on the initial conditions determined by the topological heterogeneity of the network.


\end{document}